\documentclass[graybox]{svmult}

\usepackage{mathptmx}       
\usepackage{helvet}         
\usepackage{courier}        
\usepackage{type1cm}        
\usepackage[numbers,sort&compress]{natbib}	
\usepackage{makeidx}         
\usepackage{graphicx}        
\usepackage{multicol}        
\usepackage[bottom]{footmisc}
\usepackage{amsmath,amssymb,amsfonts}
\usepackage{graphicx}
\usepackage{bm}
\usepackage{dsfont}
\usepackage{color}
\usepackage{slashed}
\usepackage[colorlinks=True,linkcolor=blue,citecolor=blue,urlcolor=blue]{hyperref}
\usepackage{hypcap}
\usepackage{comment}
\usepackage{ulem}

\newcommand{\pdag}{{\phantom{\dagger}}}

\newcommand{\mcl}[1]{\mathcal{#1}}
\newcommand{\mrm}[1]{\mathrm{#1}}
\newcommand{\mbf}[1]{\mathbf{#1}}
\newcommand{\bs}[1]{\boldsymbol{#1}}

\def\be{\begin{equation}}
\def\ee{\end{equation}}
\def\bea{\begin{eqnarray}}
\def\eea{\end{eqnarray}}
\newcommand{\bk}{{\bf k}}

\makeindex             

\begin{document}

\title*{Common and not so common high-energy theory methods for condensed matter physics}

\author{Adolfo G. Grushin}
\institute{Institut Ne{\'e}l, CNRS and Universit{\'e} Grenoble Alpes, Grenoble, France, and 
Department of Physics of the University of California Berkeley, Berkeley, CA, USA.
 \email{adolfo.grushin@neel.cnrs.fr}}
\maketitle
\abstract{This chapter is a collection of techniques, warnings, facts and ideas that are sometimes regarded as theoretical
curiosities in high-energy physics but have important consequences in condensed matter physics. 
In particular, we describe theories that have the property of having finite but undetermined 
radiative corrections that also happen to describe topological semi-metallic phases in condensed matter.
In the process, we describe typical methods in high-energy physics that illustrate the working principles to describe a given phase of matter and its response to external fields. They are based on three lectures given at the 2017 Topological Matter School, which are available on YouTube (click to see  \href{https://www.youtube.com/watch?v=ZmmOe61Ea4M}{Lecture I}, 
\href{https://www.youtube.com/watch?v=Gw6dmVPSI_g}{Lecture II}, and
\href{https://www.youtube.com/watch?v=UG6rwZdQabA}{Lecture III}).}

\section{Introduction: what this chapter is and what it is not}

Imagine you are (good) theoretical high-energy physicist and you come up with a fantastic theory: the F-theory.
As a good theorist you know that any theory that aspires to describe the universe 
has to comply with those symmetries that are verified up to experimental precision, e.g. Lorentz symmetry. 
This constraint comes with slightly less freedom to devise new testable theories, but also with a typically overlooked positive side that we will dive into: 
those same constraints save theories from apparently fatal ambiguities.

If, alternatively, you are a (good) theoretical condensed matter physicist, you have the freedom to come up with theories that violate fundamental symmetries of nature so long as you justify such effective scenario in a sufficiently realistic context. 
This freedom comes with a price; those ambiguities that high-energy theorists disposed of, 
can emerge when calculating observables, which however, should be well defined objects.
Their direct experimental relevance forces us to address them, and in doing so, sometimes we can explore 
an exotic land in between high-energy and condensed matter physics.

This chapter is a hopefully coherent and motivated compilation of different theoretical facts that deal with and, 
in the best case scenario, fix those ambiguities. 
Due to their historical context, they are not typically treated in field theory books despite that they keep being useful 
in the study of condensed matter, and very particularly topological phases.
This chapter is motivated and tailored to the study of current research in topological semimetals of different kinds, a focus that serves to emphasize
that keeping in mind these examples can prepare the reader for (a small part of) the unknown.

Finally, a disclaimer. 
Due to the short nature of this chapter I mostly use physically motivated plausibility arguments rather than formal arguments or proofs.
Along the way I will try to guide the interested reader towards the relevant formal literature as specifically as possible, but avoiding
severe computations in favor of physical intuition.
More generally, the reader is referred to the numerous reviews for details of Weyl semimetal physics and anomalies 
(e.g.~\cite{Armitage2017,Land16} and references therein) as well 
as other chapters of this volume as a back up of what is discussed here.

\section{\label{sec:LQED}Lorentz breaking field theories}

In this section we will define a simple field theory that we will use to exemplify some of the methods we will discuss.
This theory is simple but it can be used to understand a large fraction of the Weyl semimetal literature~\cite{Volovik:1999da,Klinkhamer:2005bi}.
Moreover, it has many interesting features and can be promoted, with intuitive generalizations, to describe other topological phases such as nodal 
semimetals.

\subsection{One useful field theory: Lorentz breaking QED}
Consider the following $4\times 4$ Hamiltonian in 3D momentum space spanned by the vector $\bk \in \mathcal{R}^3$
\begin{equation}
 \mcl{H}^{\mathbf{k}}_0 = \left(   
 \begin{array}{cc} 
 b_0+\bs{\sigma} \cdot (\bf{k}-\bf{b}) & m \\ 
 m & -b_0-\bs{\sigma} \cdot (\bf{k}+\bf{b})
 \end{array}  
 \right)\,.
 \label{eq:Hb}
\end{equation}
Here $\bs{\sigma}$ is the vector of Pauli matrices for a spin-1/2 degree of freedom and $b_{\mu}=(b_0,\mathbf{b})$ is a constant four vector. 
The matrices in this representation will be termed $\Gamma$ to distinguish them from the Dirac matrices below and serve to define a more compact representation of the above that reads
\be
\label{eq:HbGamma}
\mcl{H}^\mbf{k}_0 = \mathbf{k}\cdot\bs{\Gamma}+\Gamma_{5}b_0-\mathbf{b}\cdot\boldsymbol{\Gamma}^{b}+m \Gamma_0,
\ee
where $\bs{\Gamma}=\bs{\sigma}\otimes\tau_3$, $\Gamma_5=\sigma_0\otimes\tau_3$, $\boldsymbol{\Gamma}^{b}= \boldsymbol{\sigma}\otimes\tau_0$ and $\Gamma_0=\sigma_0\otimes\tau_1$, with $\sigma_0$ and $\tau_0$ being $2\times2$ identity matrices.
The Hamiltonian density Eq.~\eqref{eq:Hb} acts on a four component spinor that, for future convenience, we can write in terms of two component spinors $\Psi^{\dagger}= (\Psi^{\dagger}_R,\Psi^\dagger_L)$.
In high energy physics it is more common to use the action
\begin{equation}
\label{eq:lorentzbreakingS}
 S=\int d^4k \bar{\Psi}(\slashed{k}   - m + \gamma_5 \slashed{b} ) \Psi,
\end{equation}
where we have used the custom high-energy notation $\bar{\Psi}=\Psi^{\dagger}\gamma_0$ \footnote{This object is sometimes referred to as the Dirac adjoint. Its form is helpful to define objects that are Lorentz scalars such as $\bar{\Psi}\Psi$.} 
and Feynman's slashed notation $\slashed{k} = \gamma^{\mu}k_{\mu}$
with $k_{\mu}=(\omega,\mathbf{k})$ and $\mu=0,1,2,3$.
This notation is not strictly necessary, but it will help us connect with the high-energy literature.
Deducing Eq.~\eqref{eq:lorentzbreakingS} from Eq.~\eqref{eq:Hb} is straightforward if we have the Lagrangian density $\mcl{L}$ since $S=\int d^{4}k \mcl{L}$. 
We can use the standard relation between Lagrangian density and Hamiltonian density 
$\mcl{L}(t,\mbf{k})=\pi\dot{\Psi}_\mbf{k}-\Psi^{\dagger}_\mbf{k}\mcl{H}^0_\mbf{k}\Psi_\mbf{k}$.
Remembering that the generalized momentum in this case is  $\pi=i\Psi_\mbf{k}^{\dagger}$ and going to the frequency domain we can write
$\mcl{L}(\omega,\mbf{k})=\Psi^{\dagger}_{\mbf{k}}\gamma_0\left(\gamma_0\omega - \gamma_0\mcl{H}^{\mbf{k}}_0\right)\Psi_{\mbf{k}}$ using
the matrix multiplying $m$ in \eqref{eq:Hb}  ($\gamma_0=\sigma_0\otimes \tau_1$) which satisfies $\gamma_0^2=1$.
You can check that the Dirac matrices with our choice Eq.~\eqref{eq:Hb} are given by
\begin{equation}{\gamma^0} =
\left(\begin{array}{c c}
0 & \sigma_0 \\
\sigma_0 & 0
\end{array}\right) , \;
{\gamma^j} =
 \left(\begin{array}{c c}
0 & {\sigma^j} \\
{-\sigma^j} & 0
\end{array}\right), \; {\gamma^5} =
\left(\begin{array}{c c}
-\sigma_0 & 0 \\
0 & \sigma_0
\end{array}\right), \;
\label{eq:chiralbasis}
\end{equation}
or, alternatively, $\bs{\gamma}=i\bs{\sigma}\otimes\tau_2$, $\gamma_{0}=\Gamma_0$ and $\gamma_5=\Gamma_5$. 
Eqs.~\eqref{eq:Hb} and \eqref{eq:lorentzbreakingS} are the central objects of this chapter and contain the same information. In what follows we will use them interchangeably.

\begin{figure}[t]
 \includegraphics[width=\columnwidth]{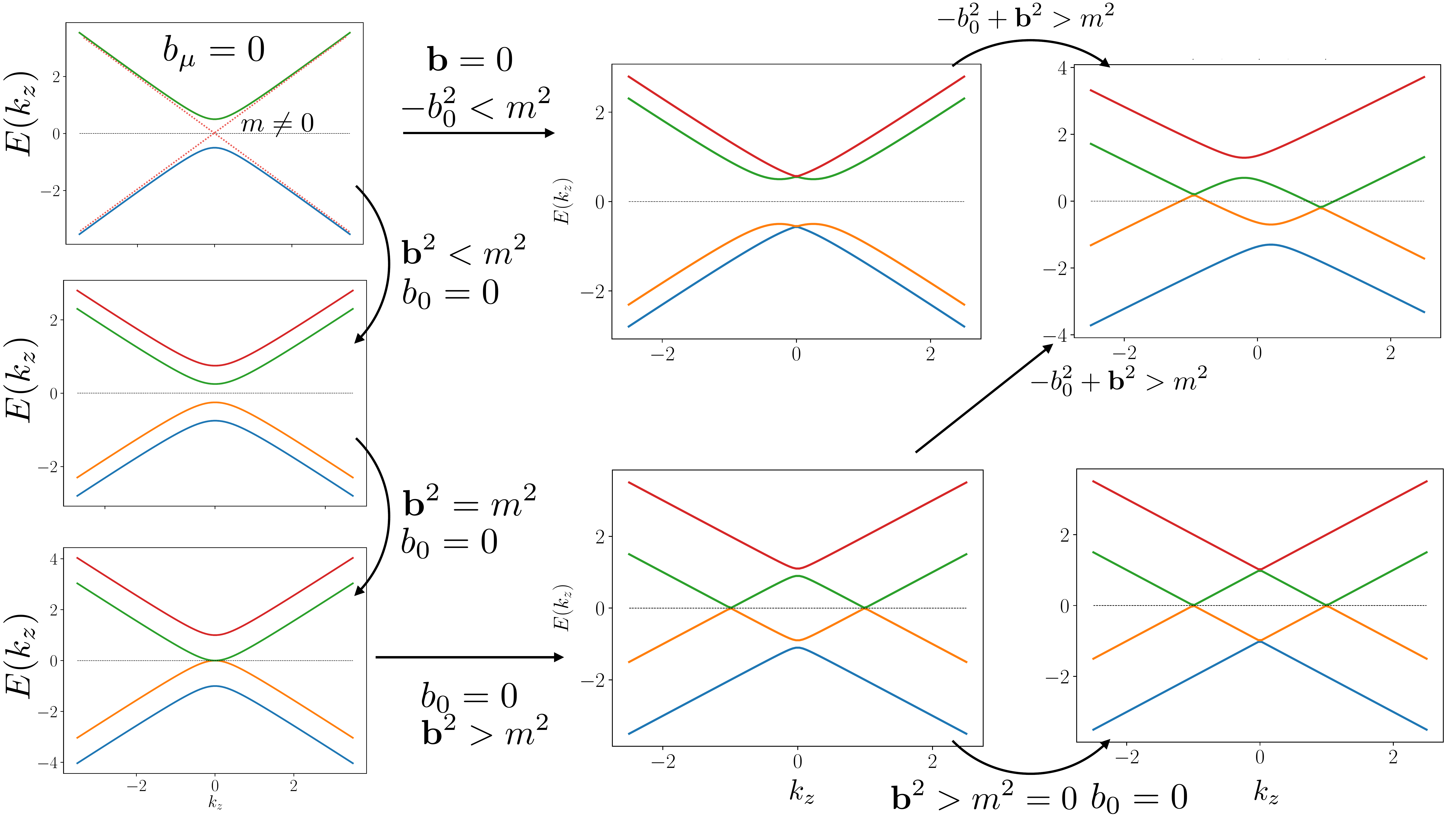}
 \caption{Band strutture of Hamiltonian given in Eq.~\eqref{eq:HbGamma} for different values of the parameters. 
 Whenever $-b^2=-b^2_0+\mathbf{b}^2<m^2$ the spectrum is gapped (see main text), while it is in semi-metallic phase otherwise.}
 \label{fig:bands}
\end{figure}
\subsubsection{Spectrum and symmetries}
Lets break down the properties of this Hamiltonian by choosing some easy limits. 
The most familiar should be the one where $b_\mu=0$ but $m\neq0$  (see Fig.~\ref{fig:bands} upper left panel).
This is the Dirac Hamiltonian where the spinor $\Psi$ satisfies the Dirac equation
\be
\label{eq:Dirac}
(\gamma_{\mu}k^{\mu}-m)\Psi = (E\gamma_0-\mathbf{k}\cdot\bs{\gamma}-m)\Psi = 0.
\ee
Solving for $E$ it is easy to find that the spectrum is gapped and has two degenerate bands $E_{\pm}=\pm\sqrt{|{\bf p}|^2+m^2}$.
The explicit form of its eigenstates can be found in any quantum field theory book (see \cite{Peskin} or, for a more condensed matter perspective, \cite{Stanescu}).
An important property of the Hamiltonian Eq.~\eqref{eq:Hb} is that it is time-reversal and inversion symmetric.
These symmetries are represented by $\mathcal{T}=i\sigma_2\otimes\tau_0$, and $\mathcal{I}=\gamma_0$, operators respectively, which 
can be explicitly be checked to commute with Eq.~\eqref{eq:Hb} with $b_\mu=0$.

Turning off the mass ($m=0$), exposes another very useful symmetry, known as chiral symmetry.
In this limit the Hamiltonian Eq.~\eqref{eq:Hb} decomposes in two $2\times2$ blocks
\be
\label{eq:HWeyl}
\mcl{H}_{\bf{k}} = \pm\bs{\sigma}\cdot \bf{k}.
\ee 
This is the Hamiltonian for two Weyl fermions $\Psi_R$ and $\Psi_L$.
The spinors $\Psi_R$ and $\Psi_L$ are known as right and left chiralities, 
and are eigenstates of $\gamma_5$ with eigenvalues $\pm1$.
The eigenvalue of $\gamma_5$ is referred to as chirality and it is a good quantum number for Weyl spinors; 
$\gamma_5$ commutes with the Hamiltonian and thus it is a symmetry. 
The chiral symmetry can be expressed as the invariance of a Hamiltonian against the continuous $U(1)$ transformation
\bea
\begin{split}
\label{eq:chiralsymmetry}
\Psi&\to& e^{\frac{i}{2}\theta\gamma_5}\Psi,\\
\bar{\Psi}&\to& \bar{\Psi} e^{\frac{i}{2}\theta\gamma_5},
\end{split}
\eea
which is a symmetry of Eq.~\eqref{eq:lorentzbreakingS} since
\be
\label{eq:comm}
\left\lbrace \gamma_{5},\gamma_{\mu}\right\rbrace = 0.
\ee
This symmetry becomes particularly explicit in the basis~\eqref{eq:chiralbasis}, which is thus known as the chiral basis.
Projecting a Dirac spinor into a Weyl spinor can be done by the projector $\mcl{P}_{\pm}=\frac{1}{2}(1\pm\gamma_5)$
\be
\label{eq:projector}
\Psi_{R/L} = \dfrac{1}{2}(1\pm\gamma_5)\Psi.
\ee
Physically, the meaning of chirality will become clearer when we couple the theory to an external electromagnetic field; for instance each chirality propagates in opposite directions when subject to a magnetic field. 
For now, we can regard this symmetry as a the mathematical statement of invariance under \eqref{eq:chiralsymmetry}.

Before moving forward, a small word of warning.
The concepts of chirality and helicity are not equivalent and often confused.
A state with definite helicity is a two spinor that is the eigenstate of the helicity operator
\be
\dfrac{1}{2}\dfrac{\bs{\sigma}\cdot {\bf p}}{|{\bf p}|}.
\ee
This operator is not, in general, a Lorentz invariant object and changes between references frames.
A massive particle with positive helicity in a given frame can be seen by another observable in a different frame with negative helicity.
Only when $m=0$ the helicity is independent of the reference frame.
In this case, it is possible to show that the states with well defined helicity have a well defined chirality as well, and the two notions coincide (see Section 7.4 in \cite{Bettini}).\\

Although both the Dirac Hamiltonian, defined by Eq.~\eqref{eq:Hb} with $b_\mu=0$ and the Weyl Hamiltonian Eq.~\eqref{eq:HWeyl} satisfies time-reversal and inversion symmetries, only the Weyl Hamiltonian posesses chiral symmetry.
From \eqref{eq:Hb} with $b_\mu=0$ but $m\neq 0$ notice that the two chiralities are coupled, ceasing to be chiral and resulting in a gapped spectrum.
Unlike time reversal or inversion symmetries, chiral symmetry is not a fundamental symmetry of any material but rather an emergent low energy symmetry.
Therefore, one should expect that $m\neq0$ in physical realizations of this Hamiltonian and thus we might conclude that a system described by low energy Weyl fermions is a very fined tuned situation.

There are in fact two possibilities to protect the Weyl fermions from gapping out due to $m$.
The first is a very physical option in condensed matter: if additional symmetries are imposed (e.g. point group symmetries) they endow the two chiralities with extra quantum numbers that we can use to impose that $m=0$ by symmetry, and the two Weyl fermions remain decoupled.
This special case is a Dirac semimetal, where the two Weyl fermions of Eq.~\eqref{eq:HWeyl} live at the same point
in the Brillouin zone but remain decoupled. 
A material that falls into the Dirac semimetal class is Na$_3$Bi~\cite{Wang2012} and corresponds to the dashed red lines in the upper left panel of Fig.~\ref{fig:bands}. 

In this chapter we will be interested in a second and richer possibility to stabilize Weyl fermions, that does not require additional symmetries.
The idea is that separating them in phase space (energy-momentum space) will effectively stabilize them, since a large momentum transfer would be needed to couple them, preventing a gap from opening.
To implement this separation we use $b_{\mu}$.
The spectrum now will depend generically on the relative size of $b_\mu b^{\mu}=b^2= b^2_0-\mathbf{b}^2$ with respect to $m^2$ (see Fig.~\ref{fig:bands}).

To analyze each case, start from a massless Dirac Hamiltonian Eq.~\eqref{eq:HWeyl}, i.e. $m=0, b_{\mu}=0$.
In this case $b^2=m^2=0$ and the theory is gapless.
If we add a small space-like $b_{\mu}=(0,\mathbf{b})$, we can diagonalize Eq.~\eqref{eq:Hb} to see that
the masless Dirac cone splits into two Weyl nodes at zero energy, that also cross at higher energies (see Fig.~\ref{fig:bands} lower right panel).
For momenta close to each Weyl node, the spectrum is still given by Eq.~\eqref{eq:HWeyl} if we measure the momentum 
relative to the Weyl node.
The Weyl node separation in this case is $\delta {\bf K}= 2 {\bf b}$.
Now add a small mass such that $-b^2>m^2$.
Such small mass hybridizes the Weyl nodes only at high-energies as shown in Fig.~\ref{fig:bands} lower central panel.
The distance between Weyl nodes in momentum space now changes to  
\be
\label{eq:weylsepmom}
\delta \mbf{K}= 2 \mathbf{b}\sqrt{1-\frac{m^2}{\mathbf{b}^2}}.
\ee
Note that, as long as $\mathbf{b}^2>m^2$ the phase is gapless and the square root is real valued.
If we keep increasing $m$ the nodes start to approach until they annihilate at $\mbf{b}^2=m^2$.
When $\mbf{b}^2<m^2$, there is a gap between all four bands, reaching the massive Dirac limit
when $\mbf{b}^2=0$. 

Adding a small $b_{0}$ does not change the basic picture (see Fig.~\eqref{fig:bands} upper central and right panels).
A finite $b_0$ will shift the Weyl nodes along the energy axis and the condition for gaplessness becomes $-b^2=-b^2_0+\mbf{b}^2>m^2$.
If this is satisfied the Weyl node separation in energy momentum space can be written compactly as 
\be
\label{eq:weylsep}
\delta K_{\mu}= 2 b_{\mu}\sqrt{1-\frac{m^2}{|b^2|}}.
\ee
With this condition, note in particular that for a time-like $b^{\mu}=(b_0,0)$, the spectrum is always gapped.
\\

%

In order to connect with physical systems it is important to note a few important symmetry properties of the Hamiltonian \eqref{eq:Hb}.
First, the spatial part $\bf{b} $ breaks time-reversal since it couples to the Hamiltonian as a 
Zeeman term $\bf{b} \cdot\bs{\sigma}$.
One can check that explicitly by applying the operator that implements time reversal symmetry defined above, $\mathcal{T}=i\sigma_2\otimes\tau_0$.
Therefore, the coupling of $\bf{b}$ can be physically regarded as a zero field magnetization which is a finite expectation value of a field~\footnote{This is a statement which is particularly evident in the Burkov-Balents model~\cite{BB11}, one of the first models of Weyl semimetals.}.

Second, the time-like part $b_0$ breaks inversion (or parity), which one can check by applying the inversion operator $\mcl{I}=\gamma_0$ to the $b_0$ term in Eq.~\eqref{eq:lorentzbreakingS}.
From Eq.~\eqref{eq:Hb} it is evident as well that it enters similar to a chirality dependent energy offset. 
This parameter can arise from inversion breaking spin-orbit coupling (e.g. see \cite{BB11}) but in general can have several physical origins to be traced back to microscopic inversion breaking perturbations.
However, it is important to note that $b_0$ is not, technically a chemical potential:
\eqref{eq:Hb} is an equilibrium Hamiltonian and $b_0$ is a parameter of it (it is observable!), unlike the chemical potential, 
which is introduced as a gauge field  (see \cite{Landsteiner:2014fw,Land16} for a discussion).

Finally and most importantly a finite $b_{\mu}$ breaks Lorentz symmetry.
Note that, since $b_{\mu}$ is a constant vector by assumption, it chooses a preferred direction in space-time and considering the above we have identified this vector as a background expectation value.
Therefore it is not allowed to transform as a Lorentz vector under Lorentz transformations.
This specific type of Lorentz transformation, the one that leave background fields invariant while changing the coordinate frame, is referred to as particle-Lorentz transformation.
It is meant to distinguish it from Lorentz frame transformations where the fields do change; for instance, a particle that experiences only magnetic field will be seen by an observer in another frame experiencing both a magnetic and an electric field.
Our theory is actually invariant under these global changes (see ~\cite{CK97} for more discussion on this issue) but is not invariant under 
particle Lorentz transformations\footnote{The difference between particle and global Lorentz transformations is simple when thinking about a particle in a box experiencing the action of gravity $\mathbf{g}$. The vector $\mathbf{g}$ sets a prefer direction, so performing a rotation, which is a transformation belonging to the Lorentz group, will leave the system invariant only if we rotate the box and the field. Rotating the box only (particle Lorentz transformation) breaks Lorentz invariance due to the fixed direction of $\mathbf{g}$.}.

\subsubsection{Coupling to electromagnetism: QED}
The above symmetry considerations, summarized in Table~\ref{table1}, combined with our previous analysis of the spectrum implies that in order to have a Weyl phase in this model
we need to satisfy two conditions: time reversal symmetry must be broken through $\bf{b} \neq 0$ and $b_{\mu}$ must be spacelike (${\bf b}^2>b_0^2$) with  $-b^2>m^2$.
If only one of the two conditions is satisfied the system with $m\neq 0$ will always be gapped.
Therefore in the theory~Eq.~\eqref{eq:Hb} a finite mass is not equivalent to being an insulator, unlike in the simple Dirac equation.

The conditions in which the Hamiltonian enters a Weyl semimetal phase, 
will have consequences when we calculate the response of a Weyl semimetal to an external electromagnetic field.
This will require introducing an external electromagnetic gauge field $A_{\mu}$ with the usual minimal (Peierls) substitution $k_\mu \to k_\mu-eA_{\mu}$ which results in
\begin{equation}
\label{eq:lorentzbreakingSA}
S[A]=\int d^4k \bar{\Psi}(\slashed{k} - m -e\slashed{A} + \gamma_5 \slashed{b} ) \Psi.
\end{equation}
This form is very suggestive: it tells us that $b_{\mu}$ couples to a Dirac fermion similarly to an electromagnetic gauge field,
but it distinguishes the two chiralities due to the presence of $\gamma_5$.
Of course this was already apparent in Eq.~\eqref{eq:Hb}.
For a high-energy theorist it is very tempting to regard $b_{\mu}$ as the chiral or axial electromagnetic field $A^{\mu}_5$ used in high-energy literature~\cite{bertlmann2000anomalies}.
However, there is an important difference: $b_{\mu}$ is itself an observable and it is a parameter in the Hamiltonian, rather than an external field.
The first issue affects our gauge freedom to change $b_{\mu}$, while the second has consequences for out-of equilibrium responses such as the chiral magnetic effect~\cite{Vazifeh:2013fe,Landsteiner:2014fw,Land16}.

The beauty of Hamiltonian~\eqref{eq:Hb} and the corresponding action ~\eqref{eq:lorentzbreakingSA} is that with a few parameters 
they capture the band structure and response of Dirac and Weyl semimetals, as well as a Dirac (trivial or topological) insulator.
In the high energy physics community this theory is known as Lorentz breaking quantum electrodynamics and has been thoroughly studied in the context of theories beyond the standard model of particle physics~\cite{CK97,CK98}.
It was recognized early on that it can describe Weyl semimetals as well, establishing a connection between these seemingly different types of systems~\cite{Volovik:1999da, Klinkhamer:2005bi, Grushin:2012cb, Zyuzin:2012ca,Goswami:2013jp}.
In the following we will take advantage of the existing high-energy field theory knowledge to infer some properties of the Weyl semimetal phase, but before doing so, we will discuss some generalizations.
\begin{table}[htp]
\caption{Summary of the symmetry properties of the different terms in Eq.~\eqref{eq:Hb}.
$\mathcal{T}$,$\mathcal{I}$ and $\Lambda$ denote time reversal, inversion and (particle) Lorentz symmetry respectively. 
When all parameters are non-zero, the Weyl node separation is set by all of them through Eq.~\eqref{eq:weylsep}. }
\begin{center}
\begin{tabular}{c|c|c|c|c|}
&$\mathcal{T}$&$\mathcal{I}$& $\Lambda$ & physical meaning\\
\hline
$m$ & yes & yes & yes & Band gap when $b_\mu =0$\\
$2\mathbf{b}$ & no & yes & no & Weyl node separation in momentum space when $m=0$. Magnetization.\\
$2b_0$& yes & no & no & Weyl node separation in energy space when $m=0$. Spin orbit coupling.
\end{tabular}
\end{center}
\label{table1}
\end{table}%
\subsection{Generalizations of Lorentz breaking field theories}

There are many interesting ways to generalize the action Eq.~\eqref{eq:lorentzbreakingSA}, anticipating its connection to condensed matter.
One quantity that has been missing, and is the first and simplest addition to the theory is the Fermi velocity.
In general the Fermi velocity will be anisotropic and so one can include its effect as a diagonal matrix $M^{\mu}_{\phantom\mu\nu}=\mrm{diag}(1,v_x,v_y,v_z)$, such that
\eqref{eq:lorentzbreakingSA} is promoted to:
\begin{equation}
\label{eq:lorentzbreakingSAM}
S[A]=\int d^4k \bar{\Psi}(\gamma_{\mu}M^{\mu}_{\phantom\mu\nu}k^{\nu} - m - e\slashed{A} + \gamma_5 \slashed{b} ) \Psi.
\end{equation}
This factor will slightly mess up the isotropy of our equations, but it is important in order to recover known lattice expressions~\cite{Grushin:2012cb}.
Fortunately, it is not unusual that when calculating response functions we can factor these out by rescaling the momenta, but this is not always true (i.e. when higher order radiative corrections are involved).
The chirality is simply the determinant of the matrix $M^{\mu}_{\phantom\mu\nu}$, since it can be shown to control the sign of the dispersion relation~\cite{Aji:2012gs}.

Additionally, note that considering an even number of copies of Eq.~\eqref{eq:lorentzbreakingSAM} with opposite values of $\mathbf{b}$ can restore time-reversal symmetry.
Recall that $\mathbf{b}$ enters like a magnetization, so if we superimpose two magnetizations with opposite directions we effectively restore time-reversal symmetry.
To again avoid the different copies from gapping out, we can separate them in momentum space, thereby breaking inversion, but preserving time-reversal symmetry.
This is in fact the case of most of the Weyl semimetals found so far, which break inversion but respect time-reversal symmetry by realizing $N>2$ pairs of Weyl nodes.
Since Eq.~\eqref{eq:lorentzbreakingSAM} can be regarded as the building block of time-reversal symmetric Weyl semimetals, we will not consider these cases here, 
although they can have richer phenomenology~\cite{Armitage2017}.

In fact, the matrix $M^{\mu}_{\phantom\mu\nu}$ is actually one of the many generalizations of QED that have been studied.
Generally one could aim to exhaust all matrices and write down all possible terms that break Lorentz invariance
using the 16 matrices in the $4\times 4$ subspace.
As we have seen, there are five Dirac matrices in $3+1$ dimensions labelled $\gamma^{\mu}$ and $\gamma_5$.
Explicitly, $\gamma_0$ is even under time-reversal and inversion while $\bs\gamma$ are odd under both since they multiply the momentum $\mbf{k}$.
The chiral matrix $\gamma_5$ is a product of all so its odd under inversion and time reversal.
To span the full space one includes the 10 matrices resulting from $\sigma^{\mu\nu}=\frac{i}{2}\left[\gamma^{\mu},\gamma^{\nu}\right]$.
Together with the identity, they span the full space of $4\times4$ matrices.
With this information we can construct a pretty general theory
\begin{equation}
\label{eq:lorentzbreakinggen}
S[A]=\int d^4k \bar{\Psi}(\tilde{\Gamma}_{\mu}k^{\mu} - \tilde{m} ) \Psi,
\end{equation}
where we have promoted $\gamma_{\mu}\to \tilde{\Gamma}^{\mu}$ such that
\bea
\tilde{\Gamma}_{\mu} &=& \gamma_{\mu} + \Gamma_{\mu}^{LV} + \Gamma_{\mu}^{CPTV},\\
\Gamma^{\mu}_{LV} &=& M^{\mu}_{\nu}\gamma^{\nu} + d^{\mu}_{\nu}\gamma_{\mu}\gamma_5,\\
\Gamma^{\mu}_{CPTV} &=& e^{\mu} + f^{\mu}\gamma_5+g^{\mu\nu\lambda}\sigma_{\nu\lambda},
\eea
and the mass term $m \to \tilde{m}$ such that 
\bea
\label{eq:genmass}
\tilde{m} = m + m_5\gamma_5+ \gamma^{\mu}a_{\mu}+ b_{\mu}\gamma^{\mu}\gamma_5+H^{nm}\sigma_{nm},
\eea
The vector $a_{\mu}$ is not very interesting, since it can be absorbed into a redefinition of the fields ($\Psi \to e^{i a_\mu x^{\mu}}\Psi$).
Many semimetals, including Weyls and nodal lines, and their phase transitions to trivial phases can be captured
only with the generalized mass term Eq.~\eqref{eq:genmass}.
For instance, a nice exercise is to compare Eq.~\eqref{eq:genmass} with the terms discussed by~\cite{BB11}.
You will notice that some of the terms in $\tilde{m}$ lead to nodal line semimetals, materials which have a gapless 1D line
node in three dimensional momentum space.
However, the Lorentz breaking generalization Eq.~\eqref{eq:lorentzbreakinggen} does not include Type-II Weyl semimetals.
These will be discussed briefly in Section \ref{sec:Beyond}.

\section{\label{sec:Lattice}Field theories on the lattice}
Quantum field theories are always effective~\cite{Georgi1993}.
This means that they are valid below or above some energy scale, that is sometimes referred to as a cut-off, be it infra-red or ultra-violet.
In condensed matter this observation is particularly important since there is always an underlying lattice that regularizes the theory after some cut-off scale.
The existence of the lattice in condensed matter naturally links with the attempt of studying gauge theories in the lattice~\cite{Rothe}.
In this section, we discuss some types of lattice generalizations for field theories.
\begin{figure}[t]
 \includegraphics[width=\columnwidth]{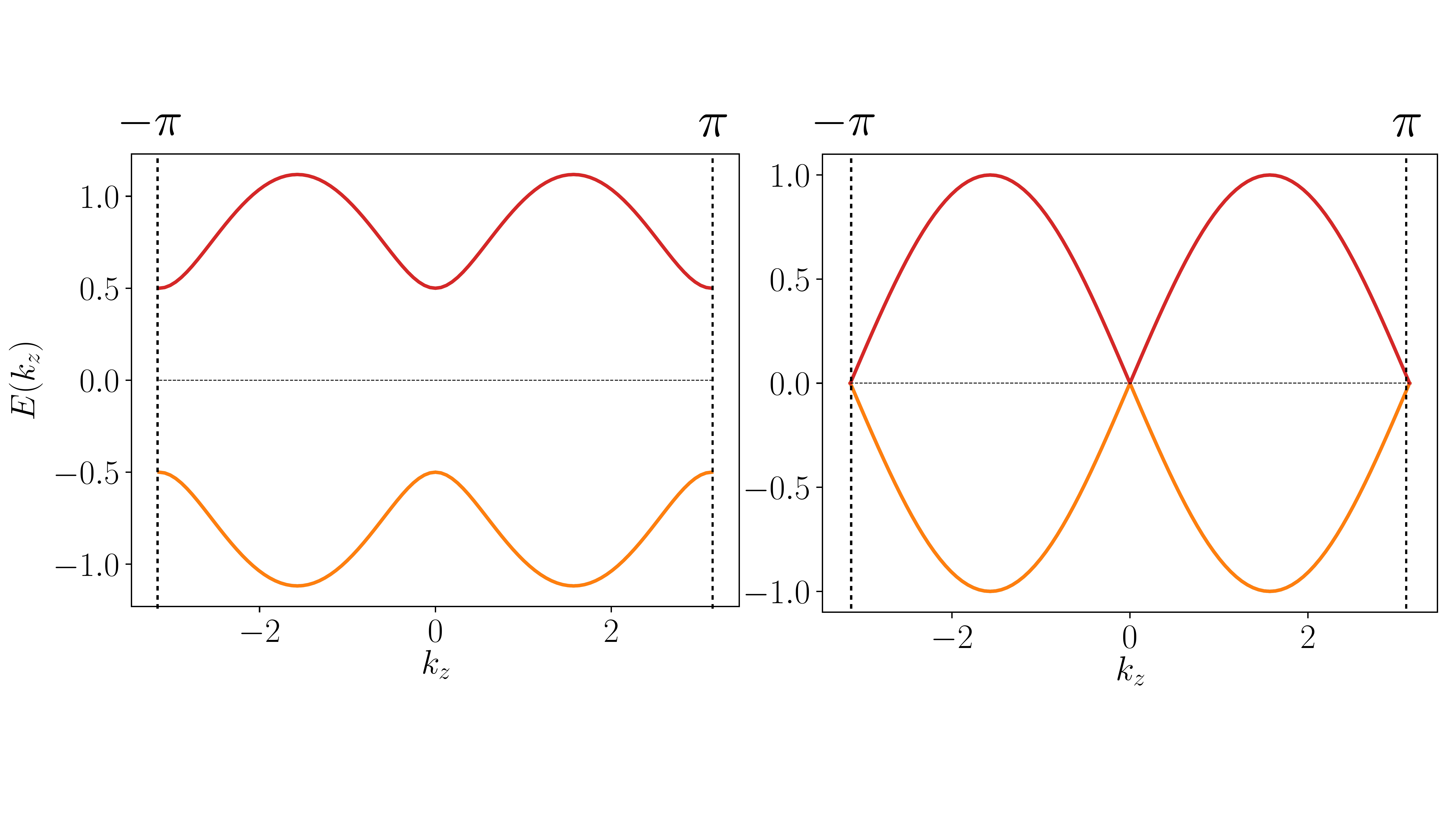}
 \caption{ Massive (left) and massless (right) ``simple" lattice fermions defined by the mapping Eq.~\eqref{eq:lattice1}  applied to Eq.~\eqref{eq:Hb} with $b^{\mu}=0$. Each band is doubly degenerate and the gap is set by $M$.
 When $M$ is zero, there are $2^d$ gapless doublers.} 
 \label{fig:bandslattice}
\end{figure}
There are many types of lattice fermions, since there are many ways of reproducing the same low energy physics from the continuum.  
Here, we will mention three different constructions that we will call ``simple" lattice fermions, Wilson fermions and Ginsparg-Wilson (or GW) fermions.
Out of the three, Wilson fermions have gained the most popularity in condensed matter, since they are the basis to understand many topological phases of matter.
Other types of lattice fermions that we will not cover include staggered fermions or twisted mass fermions (see~\cite{Rothe}).\\

\subsection{``Simple" Lattice fermions:} These are based on the most naive way of regularizing a Dirac fermion on the lattice.
They are based on the simple mapping
\bea
\begin{split}
\label{eq:lattice1}
k_i&\to \sin(k_i a),\\
m &\to M,
\end{split}
\eea
where $a$ is a lattice constant.
It is based on the intuition that close to $k_i=0$ we will recover the Dirac Hamiltonian Eq.~\eqref{eq:Hb} with $b^{\mu}=0$.
Applying this to the Dirac Hamiltonian defined by Eq.~\eqref{eq:Hb} with $b^{\mu}=0$ results in the spectrum
shown in Fig.~\ref{fig:bandslattice}.
Now try to set $M=0$.
If this mapping was to be a good description of the continuum quantum field theory at low energies,
we would like to recover one single massless Dirac fermion, which we know is invariant under chiral symmetry \eqref{eq:chiralsymmetry} so long as $M=0$.
However, close to zero energy this simple substitution leads to many copies of Dirac fermions; $2^{d}$ fermions in $d$ space dimensions (see Fig.~\ref{fig:bandslattice}).
This doubling of solutions is known as the fermion doubling 'problem'.
What this means is that applying simply Eq.~\eqref{eq:lattice1} results in massless fermions that always come in pairs, since chiral symmetry is a symmetry in the lattice.
In fact, even if we try to be smart and use the projector operator \eqref{eq:projector} to create one chiral fermion, applying to it the map~\eqref{eq:lattice1} will always result in pairs of chiral fermions with opposite chiralities.
As will be clear later on, this prevents any kind of anomaly to be present; each doubler will contribute with an opposite sign to the anomaly, since the theory on the lattice is anomaly free.
This collection of facts is known as the Nielsen Ninomiya theorem~\cite{NielNino81a,NielNino81b}. 
If can be stated as follows: if a theory is unitary, local and translational invariant there is no way to avoid the fermion doubling problem unless we sacrifice chiral symmetry in the limit $M\to0$.

\subsection{Wilson fermions:}A solution to the doubling problem where chiral symmetry is sacrificed is offered by Wilson fermions.
Wilson fermions break chiral symmetry by gapping out the doublers at the corners of the Brillouin zone with the mapping
\bea
\begin{split}
\label{eq:Wilsonfermions}
k_i &\to  \sin(k_i),\\
m &\to  M - \sum_i \sin^2(k_i/2).
\end{split}
\eea
The last term makes sure that the gap is finite irrespective of $M$ at high symmetry momentum points, except $\Gamma=(0,0,0)$ where the gap does vanish when $M=0$.
In condensed matter the last mass term is sometimes written using the identity 
$\sin^2(k_i/2) = \frac{1}{2}\left(1-\cos(k_i)\right)$.
Because of the last term in \eqref{eq:Wilsonfermions} the chiral transformation \eqref{eq:chiralsymmetry} is no longer a symmetry when $M\to0$, but the theory is free of doublers.

Wilson fermions are a constant source of inspiration for constructing models of topological phases.
A phenomenologically rich two dimensional (2D) Wilson fermion is
\be
\mcl{H}_{\mrm{CI}}= \sin(k_x)\sigma_x+ \sin(k_y)\sigma_y+\left(M-\cos(k_x)-\cos(k_y)\right)\sigma_z.
\ee
This is the simplest Chern insulator model and its main property is that it breaks time-reversal symmetry and has a finite Hall effect.
It is therefore one of the simplest topological phases (see other chapters in this volume). 
A 3D topological insulator is in fact a 3D Wilson fermion:
\bea
\label{eq:TI}
\mcl{H}_{\mrm{TI}}= \sin(k_x)\Gamma_1+ \sin(k_y)\Gamma_2 + \sin(k_z)\Gamma_3\\
+\left(M-\cos(k_x)-\cos(k_y)-\cos(k_z)\right)\Gamma_0.
\eea
where we used the $\Gamma_i$ and $\Gamma_0$ defined below Eq.~\eqref{eq:HbGamma}. 
Of course as we change dimensions, the discrete symmetry representations change, and different models respect different symmetries.
The properties of these two models, their symmetries and relation to quantum field theories can be found for example in~\cite{QHZ08}.

Now we are in place to construct a lattice generalization of the theory Eq.~\eqref{eq:Hb} using the Wilson fermion rules.
Using Eq.~\eqref{eq:Wilsonfermions} we can write our Lorenz breaking QED in the lattice as~\cite{Vazifeh:2013fe}
\be
\mcl{H}_{\mrm{WSM}} = \mcl{H}_{\mrm{TI}} + b_{i}\Gamma_{i}^{b}+b_{0}\Gamma_{5},
\ee
where $\mcl{H}_{\mrm{TI}}$ was defined by Eq.~\eqref{eq:TI} and $\Gamma^{b}_i$ and $\Gamma_5$ are defined under Eq.~\eqref{eq:HbGamma}.
This simple Hamiltonian has a very rich phase diagram including weak, strong, trivial insulators as well as Weyl semimetals
with $1,2$ or $3$ pairs of Weyl fermions~\cite{VGB15} .
It can thus very easily help to describe interfaces between topological insulating and semi-metallic phases by promoting its parameters
to be space dependent~\cite{VGB15,grushin2016inhomogeneous}.

\subsection{Ginsparg-Wilson fermions} 
There is a way of solving the fermion doubling problem less familiar in the context of condensed matter physics using a different kind of lattice fermions.  
These type of lattice fermions are known as Ginsparg-Wilson (GW) fermions, which preserve chiral symmetry up to lattice artifacts.
The exact symmetry they posses is a generalization of the symmetry \eqref{eq:chiralsymmetry} that can be written as
\bea
\label{eq:chiralsymmetrylatt}
\begin{split}
\Psi&\to& e^{\frac{i}{2}\theta\gamma_5(1-\frac{a}{2}D)}\Psi,\\
\bar{\Psi}&\to& \bar{\Psi} e^{\frac{i}{2}\theta\gamma_5(1-\frac{a}{2}D)},
\end{split}
\eea
where $a$ is the lattice constant.
They acquire this symmetry if we define the GW fermion as a type of non-local Dirac fermion
\be
\label{eq:GWfermions}
S = \sum_{x,y} \bar{\Psi}_x \left(D_{x,y} - m\delta_{x,y} \right)\Psi_y,
\ee
where $D_{x,y}$ is a non-local lattice operator that is required to satisfy the commutator relationship
\be
\label{eq:DoperatorGW}
\left\lbrace \gamma_5, D \right\rbrace = aD\gamma_5D,
\ee
that recovers Eq.~\eqref{eq:comm} when we take the limit of $a\to 0$.
This construction is quite interesting since it allows to study fermions on a lattice with chiral symmetry.
Many of the properties of massless Dirac fermions translate upon the replacement $\gamma_5\to \gamma_5(1-\frac{a}{2}D)$.
The properties match those of the continuum theory, albeit differences of order $\mcl{O}(a)$ should be expected.
A lesson to take from this is that, sometimes, corrections of order $\mcl{O}(a)$ can be crucial
to understand the linear response of a certain phase.
One specific form for the operator $D$ which is local and free of doublers was found over a decade after the GW proposal~\cite{NN93}, and is referred to as overlap fermion.
The explicit form of $D$ for overlap fermions will not be given here, but can be found easily in standard text-books \cite{Rothe}.
\\

\section{Quantum field theories can be finite but undetermined}

Jackiw, among others, noticed that some quantum field theories have
radiative corrections that are superficially divergent, but are finite 
(see~\cite{Jackiw2000} for a review, which we will follow closely in this section).
They are therefore regularization dependent, and thus ambiguous!
One could ask: Why should we care?
Anyway we could can argue that renormalizable and super-renormalizable 
field theories should be supplemented by a measurement and non-renormalizable field theories are already pathological (in a very definite sense!).
Such measurements sets a renormalization scale and gives us boundary conditions to solve
the flow equations for the coupling constant~\cite{Peskin}.
The difference here is that the constants do not necessarily flow but 
do need an experimental input.
No big deal right?

But let's step back for a moment. 
Imagine that one of these field theories actually describes low energy electrons in a material 
(or whichever degree of freedom for that matter).
It seems we would have a problem; our low energy field theory would not tell us what the values of some observables are, even if the theory is finite.
It is tempting to say that, in condensed matter, the answer is simply that the lattice fixes the regularization rendering a finite result,
which is certainly true.
As it turns out, understanding the exact way this happens gives us plenty of useful 
information about the phase this theory describes.
The kinds of field theories known so far that have this property are all tied to topological semi-metallic phases of matter that exhibit quantum anomalies and thus the focus of the following sections.

\subsection{A 1+1 D example: The Schwinger model}
\begin{figure}[t]
\includegraphics[width=\columnwidth]{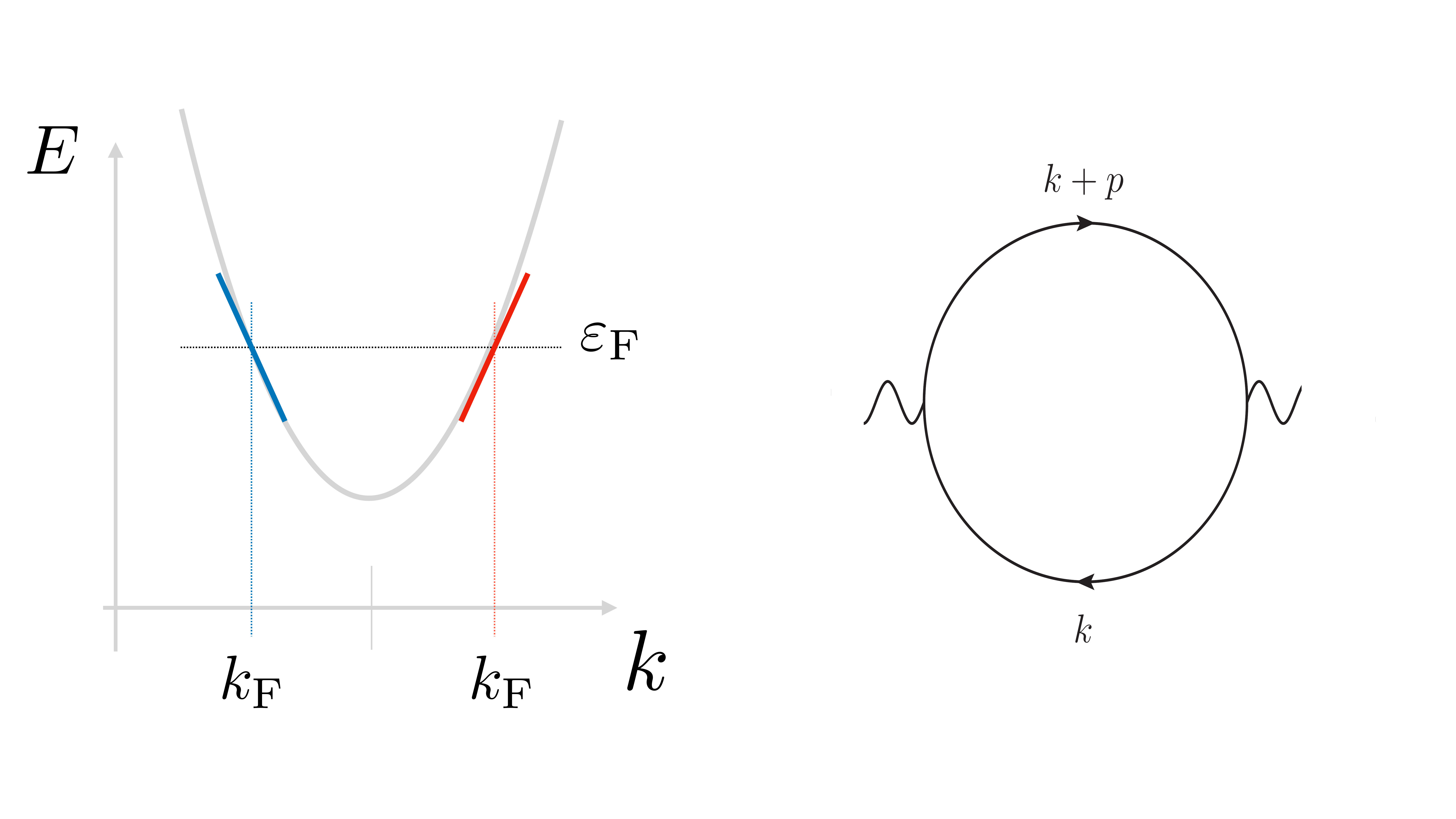}
 \caption{The left panel shows how two chiral fermions described by the action of Eq.~\eqref{eq:ch} arise from linearizing a quadratic band dispersion close to the Fermi level $\varepsilon_\mathrm{F}$. 
 The right panel shows the polarization bubble $\Pi^{\mu\nu}(p)$, where a solid line represent the Green's function.
 } 
 \label{fig:1Dfig}
\end{figure}
Let us work out a simple example first, massless QED in 1+1, or in other words, two counter propagating one dimensional chiral fermions.
It is defined by the action
\be
\label{eq:ch}
S[A]=\int d^{2}k \bar{\Psi}(\slashed{k}-e\slashed{A})\Psi,
\ee
which also defines the propagator
\be
\label{eq:gr}
G_k = \dfrac{i}{\slashed{k}}.
\ee
As per usual $\slashed{k}=k^\mu\gamma_{\mu}$ and the three necessary $\gamma$ matrices can be taken to be the three Pauli matrices $\gamma_0=\sigma_y$
$\gamma_1=i\sigma_x$ and $\gamma_5=\sigma_z$.
This is a really simple theory of two chiral modes that disperse with energy $E_k=\pm k$.
It as a linearization of a simple quadratic dispersion around the Fermi level as shown in Fig.~\ref{fig:1Dfig} left panel.
Imagine you want to find the response of this theory to an external electromagnetic field $A_{\mu}$.
You will have to calculate the expectation value of the current $j^{\mu}$ using perturbation theory in $A_{\mu}$
\be
\label{eq:current}
\left\langle  j^{\mu}(p) \right\rangle =\left\langle \dfrac{\delta S}{\delta A_{\mu}}\right\rangle = \Pi^{\mu\nu}(p)A_{\mu}+\dots,
\ee
where $\Pi^{\mu\nu}$ is the polarization function.
As described in the Appendix, the polarization function defines the effective action that governs linear response
\be
\label{eq:Seff}
S_{\mrm{eff}}[A]= \int d^4p A_{\mu}(p) \Pi^{\mu\nu}(p)A_{\nu}(-p).
\ee
The polarization function is given by
\bea
i\Pi^{\mu\nu}(p)&=&e^2\int \dfrac{d^2\; k}{(2\pi)^2}\mrm{Tr}\left[\gamma^{\mu}G_k\gamma^{\nu}G_{k+p}\right]\\
&=&e^2\int \dfrac{d^2\; k}{(2\pi)^2}\mrm{Tr}\left[\gamma^{\mu}\dfrac{i}{\slashed{k}}\gamma^{\nu}\dfrac{i}{\slashed{k}+\slashed{p}}\right].
\eea
and can be represented by the Feynman diagram in Fig.~\ref{fig:1Dfig} right panel.
In quantum field theory, the limits of integration are $\pm\infty$.
This integral is, by power counting, logarithmically divergent, and thus we need to regularize it.
Let us isolate the divergent part of the integral and evaluate the finite part.
This can be done following any of the standard quantum field theory text books (e.g. \cite{Peskin}, Chapter 7)
only noting that in 1+1 dimensions, $\mrm{Tr}[1]=2$ and that $\gamma_5\gamma^{\mu}=\epsilon^{\mu\nu}\gamma_\nu$.
Formally, we can write the result as:
\bea
\label{eq:PiSchw}
\Pi^{\mu\nu} &=& \Pi^{\mu\nu}_\infty + \Pi^{\mu\nu}_{\mrm{finite}},\\
\Pi^{\mu\nu}_{\mrm{finite}} &=& \dfrac{1}{\pi}\left(\dfrac{1}{2}g^{\mu\nu}-\dfrac{p^\mu p^\nu}{p^2} \right),\\
\Pi^{\mu\nu}_\infty &=& \dfrac{a}{2\pi} g^{\mu\nu}.
\eea
To calculate the constant $a$ we could use for instance dimensional regularization, Pauli-Villars regularization or a high-energy cut-off. 
In doing so, we would realize that they all lead to a non-divergent result; $a$ is a number, but this number depends on the regularization.
What fixes the value of $a$ is in fact the requirement that the theory should be gauge invariant.
Gauge invariance, is equivalent to charge conservation: the four-divergence of the current \eqref{eq:current} should be zero.
This implies that its four-divergence is zero, or in other words that $p_\mu\Pi^{\mu\nu}=0$.
Imposing this condition immediately sets $a=1$ and we can breathe again!\footnote{Dimensional regularization and Pauli-Villars in fact automatically give a transverse photon ($a=1$) since they both preserve gauge invariance.}

However, one key point of this section is that for certain models, the requirement of gauge invariance is not enough to fix the undetermined coefficient.
This can be illustrated by the chiral Schwinger model, defined as
\be
\label{eq:chiralSchwinger}
S[A] = \int d^2k \bar{\Psi}\left(\slashed{k}-2eP_R\slashed{A}\right)\Psi.
\ee
Remember that the projector $P_R = \frac{1}{2}(1+\gamma_5)$ projects out the right chirality fermion.
We can now ask what is the response of this model to an external field.
A similar exercise as for the Schwinger model leads to~\cite{Jackiw85}
\bea
\label{eq:PichiralSchw}
\Pi^{\mu\nu} &=& \Pi^{\mu\nu}_\infty + \Pi^{\mu\nu}_{\mrm{finite}},\\
\Pi^{\mu\nu}_{\mrm{finite}} &=& -\dfrac{1}{\pi}(g^{\mu\alpha}+\epsilon^{\mu\alpha})\dfrac{p_\alpha p_\beta}{p^2}(g^{\beta\nu}-\epsilon^{\beta\nu}), \\ 
\Pi^{\mu\nu}_\infty &=& \dfrac{a}{\pi}  g^{\mu\nu}.
\eea
Now notice that
\be
\label{eq:CA11}
p_\mu\Pi^{\mu\nu}= \dfrac{1}{\pi} \left(p^{\nu}(a-1) + p_{\mu}\epsilon^{\mu\nu}\right);
\ee
the dimensionless constant $a$ is not fixed! 
No value of $a$ sets $p_\mu\Pi^{\mu\nu}=0$.
Note that if we add up to this result the left chirality ($\epsilon^{\mu\nu}\to-\epsilon^{\mu\nu}$) we recover the complete Schwinger model calculation and gauge invariance.
A lesson we can already grasp is that theories with a single chiral fermion seem to have an inherent ambiguity to them.
This statement is of course nothing but the statement that the conservation of chiral charge is anomalous and importantly, regularization dependent.

One could argue that the arbitrariness of $a$ is not a problem.
The lattice theory always has two fermions of opposite chirality so we will always recover gauge invariance in the lattice (i.e. the Schwinger model).
Although this is in general true, note that i) the two chiral fermions can in principle be probed independently (e.g. if they are realized at different edges of a sample) and ii) even when chirality is restored, the answer can be intrinsically ambiguous, and not determined by the bulk theory, as we will see in the next subsection. 

\subsection{A 3+1 D example: Lorentz breaking QED}
In Section \ref{sec:LQED} we introduced the following Lorentz breaking QED theory
\begin{equation}
\label{eq:lorentzbreakingSAM2}
S[A]=\int d^4k\; \bar{\Psi}(\gamma_{\mu}M^{\mu}_{\phantom\mu\nu}k^{\nu} - m - e\slashed{A} + \gamma_5 \slashed{b} ) \Psi,
\end{equation}
which we argued describes a Weyl semimetal if $-b^2> m^2$.
Since experiments can probe the response of these materials to external perturbations, this section is devoted to calculating 
such response in linear order (for technical details see \cite{Grushin:2012cb}).

As in QED, the coupling to the external electromagnetic field is 
$j^{\mu}A_{\mu}$, where $j_{\mu}$ is the current operator, defined by the free fermionic action, $j_{\mu}= \frac{\delta S}{\delta A_\mu}$.
Taking care of $M^{\mu}_{\phantom\mu\nu}$ containing the Fermi velocities we have
\begin{eqnarray}
j^{\mu}=M^{\mu}_{\phantom\mu\alpha}\bar{\psi}_{\mathbf{k}}\gamma^{\alpha}\psi_{\mathbf{k}}.
\end{eqnarray}
Using Eq.~\eqref{eq:current} we can define the polarization function in linear response $\Pi^{\mu\nu}(p,b)$ which is given by
\begin{eqnarray}
\label{eq:current2}\nonumber
\left\langle j^{\mu}\right\rangle  &=& \left\langle M^{\mu}_{\hspace{2mm}\alpha}M^{\nu}_{\hspace{2mm}\beta}\bar{\psi}_{\mathbf{k}}\gamma^{\alpha}\psi_{\mathbf{k}}
\bar{\psi}_{\mathbf{k}}\gamma^{\beta}\psi_{\mathbf{k}}\right\rangle A_{\nu}\\
&=&M^{\mu}_{\hspace{2mm}\alpha}M^{\nu}_{\hspace{2mm}\beta}\Pi^{\alpha\beta} A_{\nu}.
\end{eqnarray}
The polarization function $\Pi^{\mu\nu}$ is the usual photon self-energy bubble diagram
\begin{eqnarray}
\label{eq:bubble}
\Pi^{\mu\nu}(p,b)= \dfrac{e^2}{v_{x}v_{y}v_{z}}\int \dfrac{dk^4}{(2\pi)^4}\mathrm{Tr}\left\lbrace\gamma^{\mu}G(k,b)\gamma^{\nu}G(k+p^\prime,b)\right\rbrace.
\end{eqnarray}
that contains two Green's functions, this time defined by
\begin{eqnarray}
\label{eq:Gb}
G(k,b)=\dfrac{i}{\slashed{k} -m - \slashed{b} \gamma_{5}}.
\end{eqnarray}
The velocity prefactor $1/v_{x}v_{y}v_{z}$ appears by rescaling the integrated loop momentum $k$.
This rescaling defines $p^{\prime\mu}=M^{\mu}_{\phantom\mu\nu}p^{\nu}$, the rescaled external four-momentum vector. 
Note that in this case the polarization function depends not only on the momentum $p_{\mu}$ but also on $b_{\mu}$ so it can be expressed as $\Pi^{\mu\nu}(b,p)$.
It can be separated into odd and even part with respect to the interchange $\mu \leftrightarrow \nu$.
Anticipating where the ambiguity in this theory lies, we will be interested in the odd part, which can be defined as
\be
\label{eq:Pioddb}
\Pi^{\mu\nu}_{\mrm{odd}}(b,p) = \epsilon^{\mu\nu\rho\sigma}p_{\rho}b_{\sigma}K(p,b,m),
\ee
where $K(p,b,m)$ is a scalar function and $\epsilon^{\mu\nu\rho\sigma}$ is the Levi-Civita fully antisymmetric tensor \footnote{The even part has been also calculated (see~\cite{A06}), 
but the discussion on its physical implications lies outside of the topic of this short chapter.}.
In a remarkably beautiful paper, Perez-Victoria showed how to calculate $\Pi^{\mu\nu}_{\mrm{odd}}(b,p)$ to all orders in $b$ \cite{Perez99}
\begin{eqnarray}\label{zerofreq}
\Pi^{\mu\nu}_{\mathrm{odd}}= \dfrac{e^2}{v_x v_y v_z}\epsilon^{\mu\nu\rho\sigma}p^{\prime}_{\rho}\left\{
	\begin{array}{ll}
		 C_{\sigma} &\mbox{if } -b^2 \leq m^2 \\
	    C_{\sigma}-\dfrac{b_{\sigma}}{2\pi^2}\sqrt{1-\dfrac{m^2}{b^2}}  &\mbox{if }  -b^2 \geq m^2
	\end{array}
\right. ,
\end{eqnarray} 
where $C_{\sigma}$ is an finite but undetermined constant four-vector \cite{JK99,Chung99,Perez99}.  
Introducing \eqref{zerofreq} into \eqref{eq:current2} one obtains the response of the Weyl semi-metal to an external electromagnetic field

\begin{eqnarray}\label{zerofreqcurr}
\left\langle j^{\mu}_{\mathrm{odd}}\right\rangle = \dfrac{e^2M^{\mu}_{\phantom\mu\alpha}M^{\nu}_{\phantom\mu\beta}}{v_x v_y v_z}\epsilon^{\alpha\beta\rho\sigma}p^{\prime}_{\rho}A_{\nu}\left\{
	\begin{array}{ll}
		 C_{\sigma} &\mbox{if } -b^2 \leq m^2 \\
	    C_{\sigma}-\dfrac{b_{\sigma}}{2\pi^2}\sqrt{1-\dfrac{m^2}{b^2}}  &\mbox{if }  -b^2 \geq m^2
	\end{array}
\right. ,
\end{eqnarray} 
Again we find a finite but undetermined result, which looks quite complicated.
There are several considerations that can help us digest this calculation, and come to terms with this ambiguity.
First, lets see what we can learn from the terms that do not involve $C$.
The interesting regime in this case occurs for a space-like $b_{\mu}$ such that ${\bf b}^2 > b^2_0$, which 
corresponds to a gapless theory (see Section. \ref{sec:LQED}).
A simple limiting case is $b_0=0$.
Then using that the gauge potential can be expressed as 
$A_{i}=E_{i}/\omega$ in terms of the electric field $E_{i}$, we can recognize that the spatial current is 
\be
\label{eq:Hallcont}
 \mathbf{j}\propto \sqrt{1-\dfrac{m^2}{b^2}} \mathbf{b}\times\mathbf{E}=  \delta\mathbf{ K}\times\mathbf{E},
\ee
where we recovered the Weyl node separation $\delta\mathbf{ K}$ from Eq.~\eqref{eq:weylsepmom}.
In other words, the part of the Hall conductivity that is independent of C (and thus not-ambigous) is proportional to the Weyl node separation.
This is good since therefore this calculation can recover a known result.
But then, what is the role (and the correct value!) of C? 


Of course, we should expect that a decent lattice theory has to fix $C$ in some way.
As we will now discuss, the answer is not unique, and this is quite physical.
First, lets convince ourselves that we should not be surprised that $C$ indeed can be arbitrary.
Recall that $\Pi^{\mu\nu}$ determines the effective action $S_{\mrm{eff}}[A]$ through Eq.~\eqref{eq:Seff}.
Inserting the form of the odd part of the polarization function Eq.~\eqref{eq:Pioddb} into Eq.~\eqref{eq:Seff} we can write
\be
\label{eq:CFJ}
S_{\mrm{eff}}[A]_\mrm{odd}= \int d^4p\; A_{\mu}(p)\left[\epsilon^{\mu\nu\rho\sigma}p_{\rho}b_{\sigma}K(p,b,m)\right]A_{\nu}(-p).
\ee
You may recognize this action as a Chern-Simons action, which in real space has the schematic form $\varepsilon bA\partial A$.
One might recall that Chern-Simons terms can only occur in odd space-time dimensions, which is not our case. 
It is the existence of a finite four-vector $b_\mu$ which allows us to write Eq.~\eqref{eq:CFJ}.
This type of functional form is known as the Carroll-Field-Jackiw (CFJ) term~\cite{CFJ90}
\be
\mcl{L}_\mrm{CFJ}= c_{\mu}\varepsilon^{\mu\nu\rho\sigma} F_{\nu\rho}A_{\sigma},
\ee
where $c_{\mu}$ is a constant. 
It is named after the three physicists that considered 
it as an extension of Maxwell's electrodynamics that broke Lorentz invariance.
This addition to Maxwell's equations has very interesting consequences, including, but not limited to, a Faraday effect, 
birefringence~\cite{CFJ90} or even a repulsive Casimir effect~\cite{Grushin:2012cb,WAG15}.

There are two important mutually related features of a Chern-Simons action i) it is not gauge invariant and ii) it describes a system with a Hall effect.
The latter is of course consistent with Eq.~\eqref{eq:Hallcont}.
The former gives us a hint of why it is ambiguous.
Imagine that by choosing a gauge invariant regulator we impose the gauge invariance of the Lagrangian density. 
Then the whole CFJ term is zero, since a Chern-Simons Lagrangian density is gauge non-invariant.
However, if we are less strict and choose that only the action should be gauge invariant, a term like the CFJ can survive.
The reason is that in an infinite system we hide the gauge non-invariant terms that live at the surface.
The difference between imposing gauge invariance of the Lagrangian density or the full action is equivalent
to ask whether we, through the regulator, impose gauge invariance at all momenta or only at $p^{\mu}=0$ respectively, 
since (schematically) the zero momentum Lagrangian is the real space action $\mcl{L}(q=0)= \int dx \mcl{L} = S_{\mrm{eff}}$.
Pauli Villars or dimensional regularization\footnote{It should be noted that dimensional regularization results in complications arising due to the ambiguity of 
the definition of $\gamma_5$ in odd space-time dimensions} imposes gauge invariance at all momenta and thus prohibits the appearance of the CFJ term.
Other, less strict regularizators however will allow this term to exist, since they will only impose gauge invariance at the level of the action.

Additionally, the sole fact that we are dealing with a Chern-Simons action points to the fact that any surface term can alter the value of $C$ and thus the whole term is ambiguous even ignoring the above regularization ambiguity.
This statement can in fact be proven using the path integral approach known as the Fujikawa formalism~\cite{Chung99}.
So then, how does a lattice fix $C$? 

One can make use of the fact that if $b_\mu$ is time-like, the response is completely determined by $C$.
We know from the band structure that this state is an insulator which may or may not have a Hall conductivity.
Haldane calculated the general form of the Hall conductivity in 3D~\cite{H04,H14}%
\be
\sigma^{3D}_{H,ij}= \epsilon_{ijm}\dfrac{K^m_\mrm{H}}{2\pi}\dfrac{e^2}{h},
\ee
giving an explicit expression for $\mbf{K}_{\mbf{H}}$
\be
\label{eq:KH}
\mbf{K}_\mrm{H}= \nu\mbf{G} + \sum_{i}\int_{S_i}\dfrac{\mbf{k}_F\mcl{F}}{2\pi}+\sum_{i\alpha}\int_{\partial S_i}\frac{\mbf{G}_{i\alpha}\mcl{A}}{2\pi}.
\ee
The first term is the contribution from occupied states in the Brillouin Zone $\mbf{G}$ a reciprocal lattice vector.
The last two terms encode Fermi surface contributions. 
They involve $S_i$ and $\partial S_i$ that parametrize the Fermi surface sheet $i$ and its boundary respectively.
Here $\mathbf{k}_F$ is the Fermi momentum while $\mathcal{F}$ and $\mathcal{A}$ are the Berry curvature and Berry connection respectively 
(we wont need their precise definitions here but see~\cite{H14}).

Since a time-like $b_{\mu}$ results in an insulator, the last two Fermi surface term necessarily vanish. 
For such a simple insulator, $\mbf{G}=\nu \mbf{G}_{0}$~\cite{H04} which can be interpreted as the conductivity of a stacking of 2D Hall insulators with "filling'' $\nu$ on the lattice planes stacked by $\mbf{G}_{0}$~\cite{KHW92}.
By comparing with the time-like case of Eq.~\eqref{zerofreqcurr} (i.e. the upper row), this fixes $C_{\mu}=(0,\nu\mbf{G}_0)$ \footnote{The $\mu=0$ component can be understood to be fixed to zero by the fact that there is no chiral magnetic effect in equilibrium \cite{Vazifeh:2013fe,Land14}}. 
We have found a 3D quantum Hall insulator.

The insulating phase borders a Weyl semimetal phase that is described by a space-like $b_{\mu}$.
In the simplest case where $b_0=0$, the Fermi surfaces $S_i$ are point-like and have no boundary, which excludes the last term in Eq.~\eqref{eq:KH}~\footnote{There are subtleties with this statement for finite systems due to possible non-trivial edge states, but we will not discuss them here.}.
They also have $\mathbf{k}_F=0$, so does this would mean that the second term in Eq.~\eqref{eq:KH}  is excluded and the Hall conductivity is again $\nu\mathbf{G}_0$.
However, it was noted in~\cite{H14} that $\mathbf{k}_F$ it is ambiguous under the change $\mathbf{k}_F\to \mathbf{k}_F + \mathrm{constant}$ when time-reversal is broken. 
So even if $\mathbf{k}_F=0$ the second term in Eq.~\eqref{eq:KH} has a contribution from all insulator planes perpendicular to the Weyl node separation.
This sets $\mbf{K}_\mrm{H} = 2\delta\mbf{K}e^2/h + \nu\mathbf{G}_0$. 
Thus, by comparing with the space-like case of Eq.~\eqref{zerofreqcurr} (i.e. the lower row) we can fix $C_{\sigma}=(0,\nu\mbf{G}_0)$ consistent with our previous result.
We note that there can be other equivalent ways to understand this fixing in finite systems, using the topological surface states known as Fermi arcs, that contribute to the last term in \eqref{eq:KH}, but we will not discuss that here~\cite{H14}.

To summarize, the ambiguity in the low energy theory tells us that the Weyl fermion separation should have been measured 
from a reciprocal lattice vector $b_i \to G_i-b_i$.
It is nothing but the physical result that $b_i$ is only defined modulo a lattice vector.
This is equivalent to allowing a term in the action that looks like the CFJ term
\be
\label{eq:CFJ2}
S_{\mrm{eff}}[A]_\mrm{G}= \sum_i\int d^4p\; A_{\mu}(p)\left[\epsilon^{\mu\nu\rho\sigma}G^{i}_{\sigma}p_{\rho}\right]A_{\nu}(-p),
\ee
where $G_{\mu}=(0,\mbf{G})$ where $\mbf{G}$ are  integrals of the Berry curvature 
of each disjoint group of occupied bulk bands below the Fermi level.

\subsection{Connections to the chiral anomaly}

In this section we will connect the above with the chiral anomaly, which has been thoroughly discussed both in high-energy physics and in 
condensed matter (see~\cite{Land16} for a focused review).
Without dwelling too much on the details, we will focus on its ambiguities, 
and discuss briefly how they are fixed.

\begin{figure}[t]
\includegraphics[width=\columnwidth]{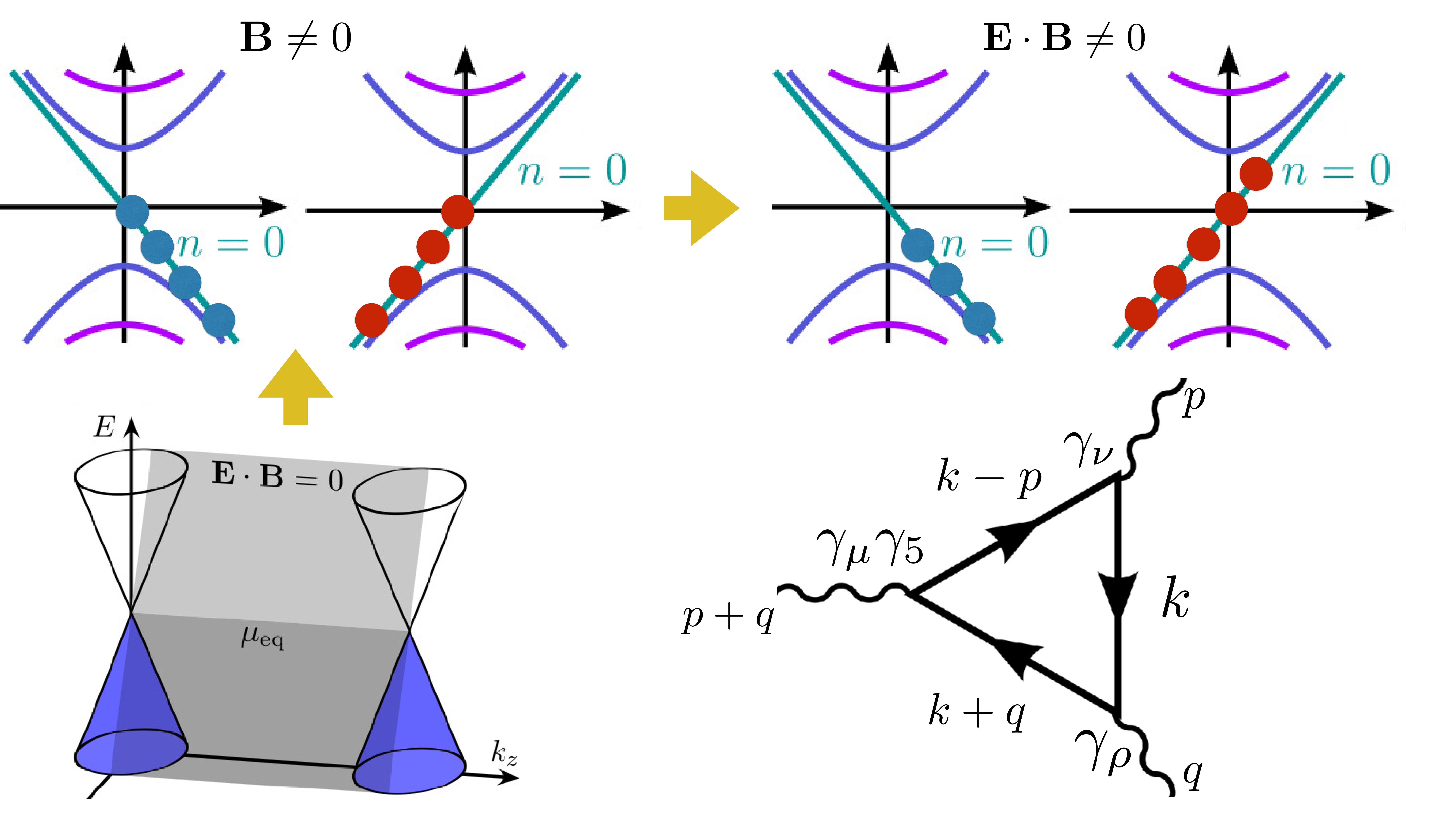}
 \caption{The magnetic field breaks the two 3+1D Weyl fermions spectrum (bottom left panel) into Landau levels which include two chiral modes dispersing along the field (top left pannel).
Applying an electric field parallel to the magnetic field (upper right panel) turns left movers (blue circles) into right movers (red circles), defining the chiral anomaly.
Diagrammatically, the chiral anomaly stems from a triangle diagram shown in the bottom right corner $\Gamma^{\mu\nu\rho}(p,q)$.
 } 
 \label{fig:CA}
\end{figure}

Consider again Eq.~\eqref{eq:bubble} and expand the Green's functions to lowest orders in $b_{\mu}$.
Using that $G(k,b)\sim G(k,0)+ iG(k,0)\gamma^{\mu}\gamma_5b_{\mu}G(k,0)$ we have that the first non-trivial order is
\bea
\nonumber
\Pi^{\mu\nu}(p,b)&\sim& e^2\int \dfrac{dk^4}{(2\pi)^4}\mathrm{Tr}\left\lbrace\gamma^{\mu}G(k)\gamma^{\nu}G(k+p)\gamma^{\alpha}\gamma_5G(k+p)\right\rbrace b_{\alpha} \\
\label{eq:bubblefirstorderb}
&+& \{ \mu \leftrightarrow \nu, \; p  \leftrightarrow -p \}\equiv \Gamma^{\mu\nu\alpha}(p,q=-p)b_\alpha.
\eea
We have identified the integrand as a triangle diagram, shown in Fig.~\ref{fig:CA} lower right panel, with a particular kinematics $\Gamma^{\mu\nu\alpha}(p,q=-p)$ with two vector vertices and one axial vertex (see for instance ~\cite{Zee2010}).
With this particular kinematic one can also isolate a divergent part of $\Gamma^{\mu\nu\rho}(p,-p)$ that depends on the regulator.
\be
\label{eq:Gammab}
 \Gamma^{\mu\nu\rho}(p,-p)_{\mrm{undet}} \sim  a\epsilon^{\mu\nu\rho\sigma}p_{\sigma},
\ee
where $a$ is finite but undetermined. 
If $b_{\mu}$ is a constant then we are safe since $p_{\mu}\Pi^{\mu\nu}=0$ and charge is conserved.
However, this conservation law is satisfied regardless of the value of $a$ so gauge invariance in fact does not fix $a$ in any way; this is the ambiguity analyzed in the previous section.

This ambiguity is in fact inherited from the chiral anomaly which is itself determined by the triangle diagram $\Gamma^{\mu\nu\rho}(p,q)$.
To understand this, lets first recall how the chiral anomaly works.
We can start with two decoupled chiral fermions in 3+1 dimensions, which from our knowledge of previous sections we can write as
\be
\label{eq:ch31}
S[A]=\int d^{4}k \bar{\Psi}(\slashed{k}-e\slashed{A})\Psi,
\ee
which is a generalization of the 1+1D action of Eq.~\eqref{eq:ch}.
To see how the chiral anomaly emerges we can follow the arguments developed in \cite{NielNino83}.
Choose $A_\mu=(A_0,\mathbf{A})$ such that ${\bf A}=B_z x {\bf e}_y$ sets a magnetic field of magnitude $B_z$ along the ${\bf e}_z$ direction, 
where ${\bf e}_{i}$ is the unit vector in $i$-direction (with $i=x,y,z$).
The Hamiltonian corresponding to Eq.~\eqref{eq:ch31} is simply the Weyl Hamiltonian Eq.~\eqref{eq:HWeyl}
that describes two Weyl fermions of chirality $\chi=\pm$ coupled to the gauge field
\begin{align}
\mcl{H}_{0}^{\chi}=\chi v_F ({\bf k}-e {\bf A})\cdot{\boldsymbol\sigma},\label{eq:weyl_b}
\end{align}
where we can set $v_\mathrm{F}=1$ for simplicity. 
Defining the magnetic length $l_B=1/\sqrt{eB_z}$ and the creation and annihilation operators
\begin{subequations}
\begin{align}
a_{k_y}^\pdag &= \frac{1}{\sqrt{2}}\left(\frac{x-k_yl_B^2}{l_B} + ik_xl_B\right),\\
a_{k_y}^\dagger &= \frac{1}{\sqrt{2}}\left(\frac{x-k_yl_B^2}{l_B} - ik_xl_B\right),
\end{align}\label{eq:bos_ll}
\end{subequations}
which obey $[{a}_{k_y}^\pdag,{a}_{k_y}^\dagger]=1$.  
we can write the Hamiltonian in the $|k_y\rangle$ basis
\begin{align}
\langle k_y|\mcl{H}_{0}^{\chi}|k_y'\rangle=\delta_{k_yk_y'}\chi v_F\,\begin{pmatrix} {k}_z&i\sqrt{2} a_{k_y}^\dagger/l_B\\-i\sqrt{2} a_{k_y}^\pdag/l_B&- {k}_z\end{pmatrix}.
\label{eq:weyl_ham_a}
\end{align}
This form of the Hamiltonian allows us to label the eigenvalues of $a^\dagger_{k_y}a_{k_y}$ by $n$, the Landau level quantum number.
The spectrum of~\eqref{eq:weyl_ham_a} comprises particle-hole symmetric bands with dispersion $E_{0,n>0}^{\chi}(k_z)= \pm\chi \sqrt{v_F^2k_z^2+2 n/l_B^2}$ and 
a chiral linearly dispersing lowest Landau level $E_{0,n=0}^{\chi}(k_z) =  \chi v_F k_z$, as illustrated in Fig.~\ref{fig:CA}.
Notice that the chiral Landau level dispersion is exactly the dispersion relation of the 1+1D field theory Eq.~\eqref{eq:ch}.
The important difference is that the bands are independent of the momentum eigenvalue $k_y$ and thus they are extremely degenerate. 
Each Landau level, including the chiral ones have degeneracy
\begin{align}
N_\text{LL}=\frac{L_x L_yB_z}{2\pi/e},
\end{align}
where $L_i$ is the length of the system in $i$-direction.
At high magnetic fields the low-energy physics is determined by the gapless lowest Landau level only.
Thus a single Weyl fermion in a strong magnetic field in the ${\bf{e}}_z$ direction is described by a macroscopically degenerate set of right- or left-moving chiral electrons with a one-dimensional dispersion $E_{0,n=0}^{\chi}(k_z)$.
Nearly without any calculation we can read off the effect of an electric field $\mathbf{E}=E_{z}\mathbf{e}_z$, set for instance by the time dependent gauge field $A_\mu=(0,0,0,E_z t )$,  on the chiral Landau levels.
Minimal substitution requires that $k_z\to k_z - eA_z = k_z - eE_zt$ and tells us that the states from two chiral branches $\pm k_z$ are created or destroyed at a rate $dk /dt = eE$ (see Fig.~\ref{fig:CA}).
If we count the charge imbalance between left and right taking into account the Landau level degeneracy we arrive to
\be
\partial_t (n_+-n_-) = N_\text{LL}\dfrac{1}{2\pi} \dfrac{dk}{dt} = \dfrac{e^2}{4\pi^2\hbar}\mathbf{E}\cdot\mathbf{B},
\ee
where we have restored $\hbar$.
This is in fact the anomalous conservation equation for the chiral current in the absence of currents, that is expressed in general as
\be
\label{eq:j5CA}
\partial_\mu j^\mu_5= \dfrac{e^2}{4\pi^2\hbar}\mathbf{E}\cdot\mathbf{B},
\ee
where $j^\mu_5=j^\mu_L-j^\mu_R$.
This result is nothing but the 3+1D generalization of the non-conservation of chiral charge Eq.~\eqref{eq:CA11}.
The total charge is conserved, but their difference is not, just as happened with Eq.~\eqref{eq:CA11}; gauge invariance is recovered when summing over chiralities.

Even thought this derivation in terms of Landau levels is physically transparent, we could have obtained this from a diagrammatic perspective which now use to connect to our previous results.
Notice that Eq.~\eqref{eq:j5CA} can be seen as arising from a Feynman diagram shown in Fig.~\ref{fig:CA}, where two legs represent the gauge fields that will compose $\mathbf{E}$ and $\mathbf{B}$ and one represents the chiral current $j^\mu_5$.
The triangle amplitude, which determines the conservation of the currents enters the vacuum expectation value of the chiral current
to second order in the external field
\be
j^{\mu}_5(l)=e^3 \int \; \dfrac{d^4q}{(2\pi)^4} \dfrac{d^4p}{(2\pi)^4} \Gamma^{\mu\nu\rho}(p,q)\delta(l-(p+q)) A_{\nu}(p)A_{\rho}(q).
\ee
A similar argument will allow us to write a contribution to $j^{\mu}$ in terms of $\Gamma^{\mu\nu\rho}$.
Thus demanding that $\Gamma^{\mu\nu\rho}(p,q)$ is transverse in all of its indices is required to conserve both currents.
This amounts to ask that  its contraction with all momenta vanishes.
However, owing to the existence of the ambiguous contribution one can show that its contractions take the form~\cite{Kaku,Land16}
\bea
 (p_{\mu}+q_\mu) \Gamma^{\mu\nu\rho}(p,q) =  \dfrac{(1+a)}{4\pi^2}\epsilon^{\nu\rho\alpha\beta}p_{\alpha}q_{\beta},\\
 p_{\nu}\Gamma^{\mu\nu\rho}(p,q) =  \dfrac{(1-a)}{8\pi^2}\epsilon^{\mu\rho\alpha\beta}p_{\alpha}q_{\beta},
\eea
where $a$ again parametrizes the regularization dependent terms.
However, unlike in Lorentz breaking QED, the ambiguity can be fixed by demanding gauge invariance, which imposes $a=1$.
In contrast, for the particular kinematics that leads to Eq.~\eqref{eq:Gammab}, gauge invariance is always a symmetry, independent of $a$.

\section{\label{sec:Beyond} Beyond Weyl fermions}

One could ask whether the above considerations can help us to study more exotic emergent fermions, 
such as Type-II (or overtilted) Weyl semimetals or three-, four-, six- and eight-fold fermions.
This section will not address this question fully, but will give two examples of what is possible.

The first example are Type-II Weyl fermions which have an over-tilted cone, such that the Fermi surface 
has a hole and an electron pocket that touch at a protected point~\cite{SGD15}.
Type-II Weyl fermions seem to break Lorentz invariance, however, they can be understood as type-I fermions in space-times which have a non-Minkowski metric, which is defined by the tilt-vector~\cite{V16,ZV16}.
A simple Hamiltonian that realizes this state is~\cite{GSZ17}
\be
\label{eq:Type-II}
\mcl{H}_{\pm}= v_{\perp}(\pm k_x\sigma_x+k_y\sigma_y)+v_z(k_z-b_z)\sigma_z+w(k_z-b_z)\sigma_0,
\ee
where $w$ is the tilt parameter.
The last term induces a time-like component to the velocity matrix $M^{\mu}_{\phantom\mu\nu}$ definied in Eq.~\eqref{eq:lorentzbreakingSAM}, 
which can be reinterpreted as a background metric.
To see this, compare Eq.~\eqref{eq:Type-II} to a Weyl fermion in curved space time
\be
\label{eq:curvedD}
\mcl{L}=\sigma^{\alpha}e^{\mu}_\alpha\partial_{\mu},
\ee
where $\sigma_{\alpha}=(\sigma_0,\bs{\sigma})$ and the tetrads $e^{\alpha}_\mu$ define the metric $g^{\mu\nu}=\eta^{ab}e^{\mu}_a e^{\nu}_b$.
This comparison leads to the definition of the line element
\be
ds^2 = g_{\mu\nu}dx^{\mu}dx^{\nu}=-dt^2+\dfrac{1}{v^2_{\perp}}(dx^2+dy^2)+\dfrac{1}{v^2_z}(dz-wdt)^2.
\ee
The tilt parameter $w$ changes the untilted spectrum ($w = 0$) to a moving reference frame with speed $w$ \cite{GSZ17}.

The second example is the collection of various other "Dirac-like" equations that describe particles beyond Weyl fermions in high energy physics.
A particularly exotic one may describe the gravitino: it is the Rarita-Schwinger Lagrangian~\cite{RS41}
\be
\mcl{L}=\dfrac{1}{2}\bar{\psi}_{\mu}(\epsilon^{\mu\rho\sigma\nu}\gamma_5\gamma_{\rho}\partial_{\sigma}-i\sigma^{\mu\nu}m)\psi_{\nu},
\ee
which describes fermions with spin-3/2. 
Recently, these type of fermions have been suggested to exist with four-fold degenerate crossings (see supplementary material in~\cite{BCW16} and the spin-3/2 fermion described in~\cite{TZZ17}).

\section{Conclusions}

In this chapter we discussed how Weyl semi-metallic phases of matter and related systems are described by ambiguous field theories, which
highlight interesting aspects of their responses to external fields. These ambiguities are connected to anomalies and are fixed by the lattice in interesting ways.
The main take-home message is that combining high energy theory literature with condensed matter phenomena can lead to interesting new insights on physical responses and
a deeper understanding of both realms of physics. It is likely that many of these techniques serve as well to understand anomalies and ambiguities in related systems such as nodal line semimetals or multi-fold fermions~\cite{Armitage2017} as well
as higher order responses which have recently shown interesting phenomenology~\cite{DeJuan2017}.

\section{Acknowledgements}

I would like to express gratitude to all collaborators and colleagues that over the years have shaped whatever you find useful in this chapter and are not to be blamed by any misstatement, 
especially Jens H. Bardarson, Jan Behrends,  Alberto Cortijo, Yago Ferreiros, Felix Flicker, Roni Ilan, Fernando de Juan, Michael Kolodrubetz, Karl Landsteiner, Titus Neupert, Sthitadhi Roy, Jorn W. F. Venderbos and Maria A. H. Vozmediano.
I am especially thankful to the organizers of the 2017 Topological Matter School in San Sebastian, Maia Vergniory, Reyes Calvo, Dario Bercioux and Jerome Cayssol, for which these lecture notes were conceived.

\appendix

\section*{Appendix: 
calculating the effective action}

This is a standard quantum field theory method ~\cite{Peskin}.
The effective action $S_{\mrm{eff}}[A]$ can be formally defined through the partition function
\be
Z[A] = e^{iS_{\mrm{eff}}[A]}\equiv \int D[\Psi]e^{iS[A]},
\ee
where we assume the action can be written as $S[A]=G_0+J^{\mu}A_{\mu}$.
In this chapter we are interested in defining $S_{\mrm{eff}}[A] $ perturbatively in $A$.
To do so we write the partition function as
\bea
Z[A] &=& \mrm{det}\left[G^{-1}_{0}-J_{\mu}A^{\mu}\right]\\
&=& \mrm{det}\left[G^{-1}_{0}\right]\mrm{det}\left[1-G_{0}J_{\mu}A^{\mu}\right].
\eea
Using that $\mrm{det}A=e^{i\mrm{Tr}\mrm{ln}A}$ and noting that the $\mrm{det}\left[G^{-1}_{0}\right]$ 
will be an overall factor that will not contribute to the calculation of observables we can define
\be
S_{\mrm{eff}}[A] = \int \dfrac{d^{d} k}{(2\pi)^d} \sum_{n=0}^{\infty} \dfrac{-1}{n} \mrm{Tr}\left[ (G_0 J_{\mu}A^{\mu})^{n} \right]. 
\ee
The second order term, responsible for linear response through \eqref{eq:current} is 
\be
S_{\mrm{eff}}[A] = \int d^{d} k A_{\mu}(p)\Pi^{\mu\nu}(p)A_{\mu}(-p) 
\ee
where~\cite{BH13} 
\be
\Pi^{\mu\nu}(p)= \int \dfrac{d^{d} k}{(2\pi)^d}\mrm{Tr}\left[G_{k-p/2} J^{\mu}_{k} G_k J^{\nu}_{k+p/2}\right].
\ee
\bibliographystyle{apsrev4-1}
%

\end{document}